\documentclass[aps,prc,showpacs,twocolumn,preprintnumbers,floatfix,
showkeys,tightenlines]{revtex4}
\usepackage{amsmath,amssymb,amsfonts}
\usepackage{graphicx}
\usepackage{color}
\allowdisplaybreaks

\newcommand{\be}{\begin{equation}}
\newcommand{\ee}{\end{equation}}
\newcommand{\bea}{\begin{eqnarray}}
\newcommand{\eea}{\end{eqnarray}}
\newcommand{\bm}{\bibitem}

\newcommand{\gm}{\gamma}

\newcommand{\ep}{\epsilon}
\newcommand{\de}{\delta}
\newcommand{\De}{\Delta}
\newcommand{\om}{\omega}

\newcommand{\lm}{\lambda}

\newcommand{\sg}{\sigma}

\newcommand{\ze}{\zeta}

\newcommand{\cz}{{\cal Z}}

\begin{document}

\setcounter{page}{1}

\vspace*{0.5 true in}

\title{Nuclear condensation and symmetry energy of dilute nuclear matter:\\
       an $S$-matrix approach}

\author{J.N. \surname{De}}
\email{jn.de@saha.ac.in}
\author{S.K. \surname{Samaddar}}
\email{santosh.samaddar@saha.ac.in}
\affiliation{Theory Division, Saha Institute of Nuclear Physics, 1/AF 
 Bidhannagar, Kolkata 700064.}  

%\date{\today} 

\begin{abstract} 
Based on the general analysis of the grand canonical partition function
in the $S$-matrix framework, the calculated results on symmetry energy,
free energy and entropy of dilute warm nuclear matter are presented.
At a given temperature and density, the symmetry energy or symmetry free 
energy of the clusterized nuclear matter in the $S$-matrix formulation
deviates, particularly at low temperature and relatively
higher density, in a subtle way,
from the linear dependence on the square of the isospin
asymmetry parameter $X=(\rho_n - \rho_p)/(\rho_n + \rho_p)$, 
contrary to 
those obtained for homogeneous
nucleonic matter. The symmetry coefficients, in conventional definition, can
then be even negative. The symmetry entropy similarly shows a very
different behavior. 
\end{abstract}


\maketitle

\section{Introduction}

Understanding properties of symmetry energy for low density
nuclear matter is of much topical
interest in both astrophysical and laboratory context. 
The supernova simulation dynamics has a sensitive
dependence  on the symmetry energy
\cite{ste}; a higher symmetry energy, for example, leads to
a lower electron ($e^-$)-capture rate in the supernova collapse
phase that may result in a stronger explosive shock. The
variations in the $e^-$-capture rate also produces changes in the
neutrino luminosities that are potentially observable. 
The isotopic abundance of relatively heavier elements
in explosive nucleosynthesis is further
directly correlated to the symmetry energy. The neutron
skin thickness of heavy nuclei has also a direct dependence on the density
variation of the symmetry energy \cite{bro,hor}. At subnormal densities
(0.2 $\le \rho /\rho_0 \le $ 1.0, $\rho_0 $ is the saturation density
of nuclear matter), 
the symmetry coefficients that are a 
measure of the symmetry energy per baryon ($e_{sym}$)
have recently been estimated from analysis
of data related to isotopic distributions \cite{bot}, isospin diffusion
\cite{tsa,che}, and isoscaling \cite{sou,she} in heavy ion reactions. 
The different analyzes give
somewhat different results. In the density domain mentioned, the symmetry
coefficient $C_E$ 
~(=$e_{sym}/X^2$)
varies with density as $C_E(\rho ) \sim 
C_E(\rho_0 ) (\rho / \rho_0 )^\gm $ with $\gm $ lying
in a broad range \cite{she,li}, between
0.55 $-$ 1.05. A more definitive answer to the question of the
density dependence of symmetry energy is still wanting.

On the theoretical side,
symmetry energy of infinite nuclear matter has been calculated over a
wide density range in many guises; many-body theories using various
nucleon-nucleon interactions or interaction Lagrangians \cite{li} have
lead to varying results \cite{fuc}. There have also been some recent
investigations on the symmetry free energy of hot nuclear matter
\cite{xu} and also of finite nuclei \cite{sam}. These calculations
differ in details, but the general qualitative behavior
of the symmetry coefficients with density do not deviate much 
from the experimental trend. A similar power law variation is exhibited;
the exponent $\gm $ 
lies mostly within the broad limits as extracted 
from experimental analyzes.

The above calculations have been done in the mean-field (MF)
approximation; the system is taken to be homogeneous nucleonic
matter. For dilute nuclear matter, however, the system 
minimizes its total energy or free energy by forming clusters
\cite{fri,pei}. A detailed knowledge of the composition of
nuclear matter is then needed to appreciate how the symmetry
energies are affected when matter gets clusterized. This
has  a direct role in a better understanding of
neutrino-driven energy transfer in inhomogeneous 
supernova matter \cite{jan}.
Using the virial expansion technique, clusterization in dilute nuclear 
matter and its import in the evaluation of the symmetry coefficients
has been investigated recently by Horowitz and Schwenk \cite{hor1}
where they considered the matter to be composed of 
$n,p,\alpha $. The investigations with the same cluster species
have been followed further \cite{kow} to connect experimentally
the temperature and density-dependent entropic contributions
to the symmetry free energy coefficient.
The resulting symmetry coefficients are found to be 
considerably larger than the corresponding ones obtained from MF
calculations. This is an important result, it shows the strong
role of clusterization and naturally calls for a realization of symmetry
coefficients if all possible permissible clusters are incorporated
in the calculation. The present article is an attempt in
this direction.

\section{Elements of theory}

The grand canonical partition function in the $S$-matrix formalism
of statistical mechanics proposed by Dashen {\it et al} \cite{das}
sets the logical framework of our calculation. The details of the 
formalism, as applied to nuclear matter is given in Ref. \cite{mal};
for the sake of completeness, the essentials are presented in the
following.

The partition function $\cz $ for the two-component nuclear
matter composed of the elementary species neutrons and protons
is written as 
\be
\cz =Tr\, e^{-\beta (H-\mu_p\hat{N}_p -\mu_n\hat{N}_n)}\,,
\ee
where $\beta $ is the inverse of temperature $T$ 
of the system, $H$ the total
Hamiltonian, $\hat{N}_{p,n}$ the number operators for protons and neutrons,
and $\mu_{p,n}$ are the corresponding chemical potentials.
 The partition function can be decomposed as
\be
\cz = \sum_{Z,N=0}^\infty \ze_p^Z \ze_n^N Tr_{Z,N}\, e^{-\beta H}\,,
\ee
where the fugacities are given by
$\ze_p=e^{\beta\mu_p},~~ \ze_n=e^{\beta \mu_n}$.
The trace $Tr_{Z,N}$ is taken over states of
$Z$ protons and $N$ neutrons. For small $\ze_p$ and $\ze_n$, the
quantity ln$\cz$ can be expanded in a virial series 
\be
\ln \cz = {\sum_{Z,N}}^{\prime} D_{Z,N}\ze_p^Z \ze_n^N \,.
\ee   
The prime on $\Sigma$ indicates that the term with $Z=N$=0 is excluded.
Evaluation of the virial coefficients $D_{Z,N}$ gives the partition function
and thence the thermodynamic behavior of the system.

Following the temperature-Green's-function method, it was shown in 
Ref.~\cite{das} that all the dynamical information concerning the
microscopic interaction in the grand potential of the system can be
collected in two types of terms so that the partition function is 
written as
\be
\ln \cz =\ln {\cz}_{part}^{(0)} + \ln {\cz}_{scat}.
\ee
The first term corresponds to contributions from stable single particle
states of clusters of different sizes 
(neutrons and protons included) formed in the infinite system
behaving like an ideal quantum gas [the superscript (0) indicates this behavior]
and the second term corresponds to contribution from multiparticle
scattering states, respectively. The particle piece can further be split
in contributions from ground states and excited states of the bound nucleon
clusters, so that
\be 
\ln {\cz}_{part}^{(0)}=\ln {\cz}_{gr}^{(0)} +\ln {\cz}_{ex}^{(0)} .
\ee 
The first term in Eq.~(5) is a sum of ideal gas terms, one for each of 
the ground states of all the possible species of mass $A$ with $Z$ protons
and $N$ neutrons that can be formed in the system,
\bea
&&\ln {\cz}_{gr}^{(0)}  =  \mp V \sum_{Z,N}g \int\!
\frac{d^3p}{(2\pi)^3}\,\times \nonumber \\ 
 && ~~~~~~~\ln \left( 1\mp \ze_{Z,N}
e^{-\beta(p^2/2Am )} \right) .
\eea
Here $m$ is the nucleon mass, {\bf p} is the momentum of the nucleus,
and $\ze_{Z,N}$ the 'effective fugacity' given by
$\zeta_{Z,N}=e^{\beta (\mu_{Z,N} +B_{Z,N})}$.
$B_{Z,N}$ is the binding energy of the nucleus 
and $\mu_{Z,N}$ its chemical potential.
From the condition of chemical equilibrium among the different species,
$\mu_{Z,N} = Z\mu_p + N\mu_n $.
 
The $\mp$ sign in Eq.~(6) corresponds to nuclei with $A$ even or odd,
obeying Bose or Fermi statistics, $V$ is the volume of the system and
$g$ is the spin degeneracy. The sum includes the original elementary
species, namely the neutrons and protons. The Coulomb interaction is
assumed absent in nuclear matter, ideally the sum in Eq.~(6) would
involve infinite terms as the maximum cluster mass $A$ can then even be
infinite. However, for applications to 
real physical systems such as neutron
star matter, Coulomb effect is to be included in the binding energies of
nuclei. In that case, the sum is finite, conditioned by the stability
of nuclei with inclusion of Coulomb in the binding energies. 
Calculations in the $S$-matrix formalism with inclusion of Coulomb
in the fragment binding energies would henceforth be referred to as
SMF, those without Coulomb would be called SNC.
Eq.~(6) can be readily expanded in a virial
series as 
\bea
&&\!\!\!\!\!\!\!\!\ln {\cz}_{gr}^{(0)} = V\sum_{Z,N}\frac{g}{\lm^3(Am)}\left(\zeta_{Z,N} \pm
\frac{\zeta^2_{Z,N}}{2^{5/2}} +\cdots\right)\,,
\eea
in powers of effective fugacities provided $|\ze_{Z,N}|<1$.
Here, $\lambda (Am)=~\sqrt{2\pi/(AmT)}$ is the thermal
wavelength of a cluster of mass $Am$. We work in natural units
$\hbar =c=1$. A nucleus in a particular excited state is taken as
a distinctly different species and can be treated in the same footing
as the ground state. The density of states is quite high in relatively
heavy nuclei, increases nearly exponentially with the square root
of excitation energy $E$ and so the contribution of the excited
states of a single nucleus is written as an integral over $E$
of the ideal gas term weighted with the level density $\omega (A,E)$,
\bea 
&&\ln {\cz}_{ex}^{(0)}=\mp V{\sum_{Z,N}}^{\prime}g\int_{E_0}^{E_s} \!\!dE~\om(A,E) 
\times\nonumber\\
&&\int\!\frac{d^3p}{(2\pi)^3}\, \ln \left( 1\mp \ze_{Z,N}
e^{-\beta(p^2/2Am +E)} \right).
\eea
For the level density, we take the expression \cite{boh}
\be
\om (A,E)=\frac{\sqrt{\pi}}{12a^{1/4}}\frac{e^{2\sqrt{aE}}}{E^{5/4}}.
\ee
The level density parameter $a$ is taken as $ A/8 $
MeV$^{-1}$, its empirical value. 

In Eq.~(8), the prime on $\Sigma $ denotes exclusion of the light nuclei
($A\le 8$) from the sum. For these nuclei, we take only the ground states;
their degeneracy factor $g$ is taken from experiments. 
For other nuclei, for both ground and excited states, $g$ is taken as
1 or 2, according as they are bosonic or fermionic. The lower limit
$E_0$ is dictated by the location of the first excited state. The 
upper limit $E_s$ is the separation energy. We take $E_0$=2 MeV and
$E_s$ =8 MeV. 

%%%%%%%%%%%%%%%%%%%%%%%%%%

The scattering piece $\ln {\cz}_{scat}$ of Eq.~(4) can be formally,
but explicitly written \cite{mal} for nuclear matter:
\bea
\ln {\cz}_{scat}&&\!\!\!\!\!\!\!\! = V\!\sum_{Z_t,N_t}\frac{e^{\beta \mu_{Z_t,N_t}}}
{\lm^3(A_tm)}\sum_{\sg} e^{\beta B_{Z_t,N_t,\sg}} \times \nonumber \\
&&\!\!\!\!\!\!\int_0^{\infty}\!\!\!d\ep~e^{-\beta\ep} \frac{1}{2\pi i} 
Tr_{Z_t,N_t,\sg}
\left({\cal A}S^{-1}(\ep)\frac{\partial}{\partial\ep}S(\ep)\right)_c .
\eea

Here, the double sum refers to the sum over all possible scattering channels,
each having its chemical potential $\mu $ and formed by taking any number
of particles from any of the stable species (proton, neutron and nuclei
in their ground and excited states) and the trace is over all plane 
wave states for each of these channels. $S$ is the scattering operator and 
${\cal A }$ the boson symmetrization or fermion antisymmetrization
operator. The subscript $c$ denotes only the connected 
parts in the diagrammatics of the
expression in the parenthesis. A channel in the set has a total number
of $Z_t$ protons and $N_t$ neutrons ($A_t~=N_t+Z_t$); $\sigma $ denotes
all other labels required to fix a channel within this set. 
$B_{Z_t,N_t,\sigma }$ is the sum of the individual binding energies of all
the particles in the channel and $\ep $ is the total kinetic
energy in the c.m. frame of the scattering partners. 
Examination of Eq.~(10) shows that channels with
larger binding energies are more important, because of the 
factor $e^{\beta B_{Z_t,N_t,\sigma }} $. Furthermore, two-particle channels
are expected to dominate over multiparticle channels with the same $Z_t$ and
$N_t$ from binding energy consideration. We therefore consider only
two-particle scattering channels.
It is convenient to divide
the channels into light ones, consisting of low mass particles 
($A < 8$, say) and heavy ones 
($A \ge 8 $), so that we write 
\be
\ln {\cz}_{scat} = \ln {\cz}_{light} + \ln {\cz}_{heavy},
\ee
as the sum of contributions from the light and heavy channels.

Experimentally it is known that the scattering of relatively heavier
nuclei is dominated by a multitude of resonances near the threshold.
The $S$-matrix elements are then approximated by resonances. Following
\cite{das1,das2}, each of these resonances are treated like an ideal gas term
and then $\ln {\cz}_{heavy}$ can be written in the same form of 
$\ln {\cz}_{ex}^{(0)}$, assuming their level density to be the
same as those of the excited states.
The integration over $E$ in  Eq.~(8) now extends from $E_s$ to 
$E_r$, the limit of resonance domination.
The damping of the integral in Eq.~(10) due
to the presence of the Boltzmann factor assures contribution only from
low energies; we therefore take $E_r \simeq $12 MeV.

The scattering channels $NN, Nt, NHe^3, N\alpha$ ($N$ refers to nucleon) 
and $\alpha \alpha$ are considered 
for evaluation of ${\cz}_{light}$. Inclusion of other light particle
scattering channels may have some influence on the present results.
In intermediate energy heavy ion collisions, for example, deuteron usually has
considerable multiplicity and the contribution to $\ln {\cz}_{light}$
from deuteron scattering channels is worth further exploration. 
With the choice of the scattering channels as mentioned
above, we then write 
\be
\ln {\cz}_{light}~=~ \ln {\cz}_{NN}+\ln {\cz}_{Nt}+\ln {\cz}_{NHe^3}
+\ln {\cz}_{N\alpha}+\ln {\cz}_{\alpha \alpha}.
\ee
We explicitly write the contribution from the $NN$ channel. If we
consider only elastic two-body scattering, the trace in Eq.~(10)
becomes a sum over the derivative of the phase shifts of the
appropriate partial waves. It gives formulas of the same form
as derived by Beth and Uhlenbeck \cite{bet} for the second virial 
coefficient. It is given as
\bea
\ln {\cz}_{N\!N}=&&\frac{V}{\lm^3(2m)} \{ (\ze_p^2+\ze_n^2)\De^{I=1}_{N\!N}\nonumber\\
&&+\ze_p\ze_n ( -3 +\De^{I=1}_{N\!N} + \De^{I=0}_{N\!N})\}.
\eea
Here $I$ is the isospin index and
\be
\Delta^I_{N\!N}=\frac{1}{\pi T}\int_0^\infty d\ep e^{-\beta \ep}
\sum_{S,L,J} (2J+1)\, \de^{NN}_{{2S+1}_{\textstyle L_{J}}}(\ep)\,.
\ee
The quantity $ \de^{NN}_{{2S+1}_{\textstyle L_{J}}}(\ep)$ refers to the NN phase
shift in the $LSJ$ channel. The contributing partial waves are determined
by $I$ through the requirement of antisymmetry on the total wave function
of the $NN$ system.
The other terms in Eq.~(12) have nearly similar forms \cite{mal}. The $\Delta$'s
for the $NN, N\alpha$ and $\alpha \alpha$ channels are evaluated in \cite{hor1}
and those for $NHe^3$ and $Nt$ are available in \cite{con}. This completes
the evaluation of partition function for nuclear matter. It is then 
straightforward to get the relevant observables like the pressure $P$,
the number density of the $i$-th species $\rho_i $, 
free energy per baryon $f$ or
the entropy per baryon $s$ from the relations,
\bea
P  =  T \frac{\ln {\cz}}{V}\,, 
~~~\rho_i =\ze_i\left ( \frac{\partial}{\partial \ze_i}
\frac{\ln {\cz}}{V}\right )_{V,T}, \nonumber \\
 f  =  \frac{1}{\rho} \left (\sum_i \mu_i \rho_i -P \right ),~~~~s=\frac{1}{\rho }
\left (\frac{\partial P}{\partial T}\right )_{\mu_i} \,, 
\eea
with the baryon density $\rho = \sum_i A_i\rho_i$.

\section{Results and discussions}

We have calculated the nuclear equation of state (EOS), fragment distributions,
the symmetry free energy, symmetry entropy and the symmetry coefficients
(symmetry energy coefficient $C_E$ and the symmetry free energy coefficient
$C_F$) for dilute nuclear matter. The calculations have been done in the
SMF and the SNC model 
where all possible nuclear species are considered and
compared, in order to explore the role of heavy species, with calculations
\cite{hor1,con} where only the light species ($n,p,d,t,He^3$ and $\alpha$) 
are taken. The latter model is subsequently referred to as 
the light species model (LSM).  The three models (SMF, SNC and
LSM) would be collectively referred to as the condensation models. To
highlight the importance of clusterization on the physical observables,
the MF results are also presented.

In practice, an asymptotic wave function may not have
a precise meaning at relatively high density; it would then be
difficult to have a meaningful expression of the partition function
in terms of the $S$-matrix elements. We, therefore, restrict our
calculations to low density nuclear matter and have considered
up to a baryon density $\rho $=0.01 fm$^{-3}$.

In Fig.~1, the EOS ($P-\rho $) is displayed for symmetric nuclear
matter ($\rho_n =\rho_p $) at  temperatures $T=$2, 4, and 8 MeV.
The dotted lines refer to calculations in the MF
model with the SkM$^*$ interaction. At low temperatures, it is seen 
that the system enters the unphysical region ($dP/d\rho < 0$)
in the density range considered. 
This can be avoided by applying Maxwell's construction.
In the SMF, because of the many-body correlations (condensation),
this unphysical behavior does not arise.
 Since Coulomb interaction is absent in nuclear matter,
to compare results from the mean-field, a set of $S$-matrix calculations
have been done with Coulomb switched off in the nuclear binding energies
(SNC). The SNC calculations are represented in the figure as 
dot-dashed lines. With isothermal compression, at lower temperatures,
the pressure levels off at very low densities as shown by the
dot-dash lines in Fig~.1(a) and 1(b), signaling a behavior like
a first-order phase transition.
At the higher temperature 8 MeV, 
the said transition
starts at a density beyond 0.01 fm$^{-3}$, it is not seen
in the figure. The sum in Eq.~(3) for the SNC calculation 
runs up to infinity in principle;  in practice, one
takes a finite sum for calculational facilitation.
We have taken the maximum mass $A_{max}$=1000. 
The results are found to be not very sensitive 
to further increase of $A_{max}$ \cite{de}. The binding energies of these
nuclei are obtained using a simple liquid-drop mass formula \cite{mye1} 
with Coulomb switched off.

To help comparison with physical systems 
such as neutron star matter, we have also considered phenomenological
binding energies (Coulomb included) \cite{mye} of the nuclei that limits
the number of terms in the  sum 
to $\sim $9000 nuclei in their ground states with $A_{max}$= 339 and
$Z_{max}$ =136. 
The EOS in the SMF are shown by the full lines.
It is seen that with Coulomb in the binding energies, the signature of
the first-order phase transition is washed out 
with monotonic increase in pressure on
isothermal compression. The results in the LSM model
are shown by the dashed lines. At a given density, in the SMF,
the fragment multiplicity is comparatively lesser as more

\newpage

\vspace*{-1cm}
\begin{figure}[b]
\includegraphics[width=0.45\textwidth,clip=false]{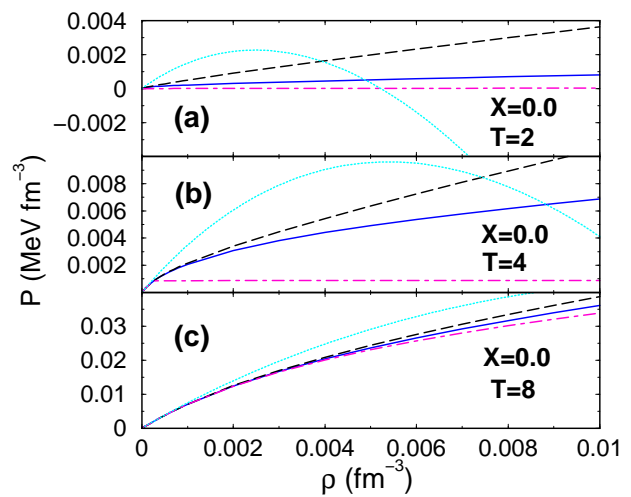}
\caption{ (Color online) The EOS for symmetric nuclear matter 
at $T$= 2, 4, and 8 MeV. Calculations are shown for models in
mean-field (dotted cyan), LSM (dashed black),
SNC (dot-dashed  magenta), and
SMF (full line, blue)} 
\end{figure}

\begin{figure}[t]
\includegraphics[width=0.45\textwidth,clip=false]{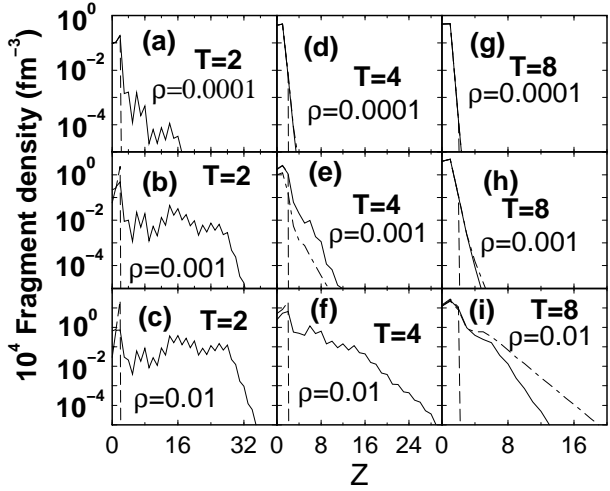}
 \caption{ The charge distributions at different temperatures and
densities as shown are compared in the models of LSM (dashed line),
SNC (dot-dashed) and SMF (full line).}
\end{figure}
\begin{figure}[b]
\includegraphics[width=0.45\textwidth,clip=false]{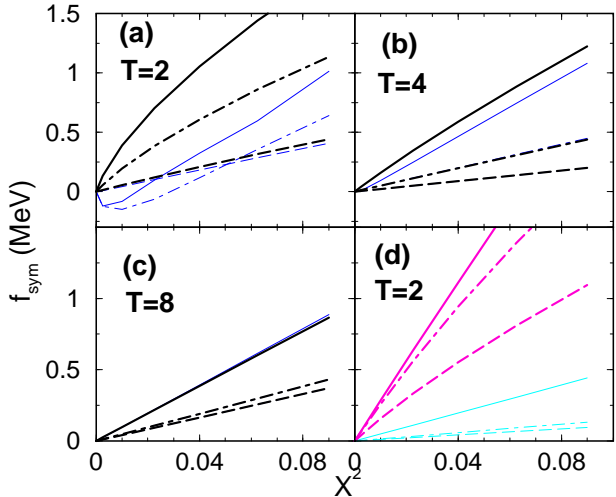}
\caption{ ( Color on-line) The symmetry free energy shown as a function
of $X^2$ at different temperatures. The dashed, dot-dashed and full 
lines correspond to $\rho=$0.0001, 0.001 and 0.01 fm$^{-3}$, respectively.
The results are calculated in the models of the mean-field (thin cyan),
LSM (thick black), SNC (thick magenta), and 
the SMF (thin blue).}
\end{figure}

\begin{figure}[t]
\includegraphics[width=0.45\textwidth,clip=false]{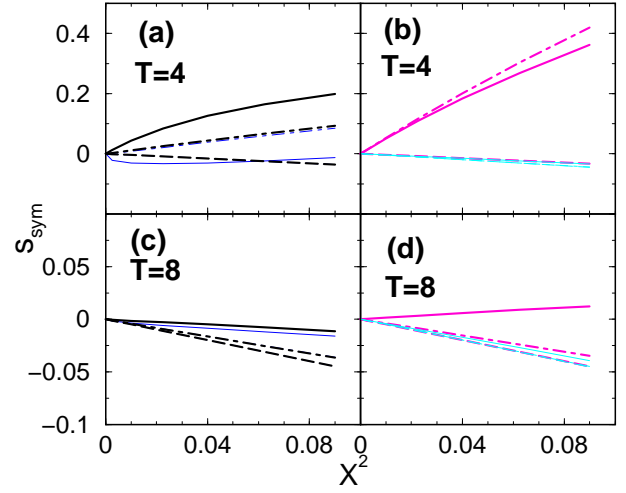}
\caption{ (Color online) The symmetry entropy $s_{sym}$ as a function
of $X^2$ at different temperatures and densities. The notations are the same
as described in the caption to Fig.~3.}
\end{figure}

\begin{figure}[b]
\includegraphics[width=0.45\textwidth,clip=false]{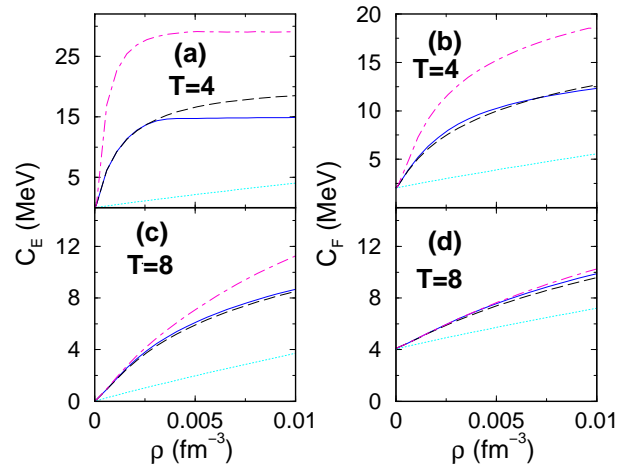}
\caption{ (Color online) The symmetry coefficients $C_E$ and $C_F$ as
a function of density at different temperatures in the models of
mean-field (dotted cyan), LSM (dashed black), 
SNC (dot-dashed magenta) and SMF (full lines, blue).}
\end{figure}

\begin{figure}[t]
\includegraphics[width=0.45\textwidth,clip=false]{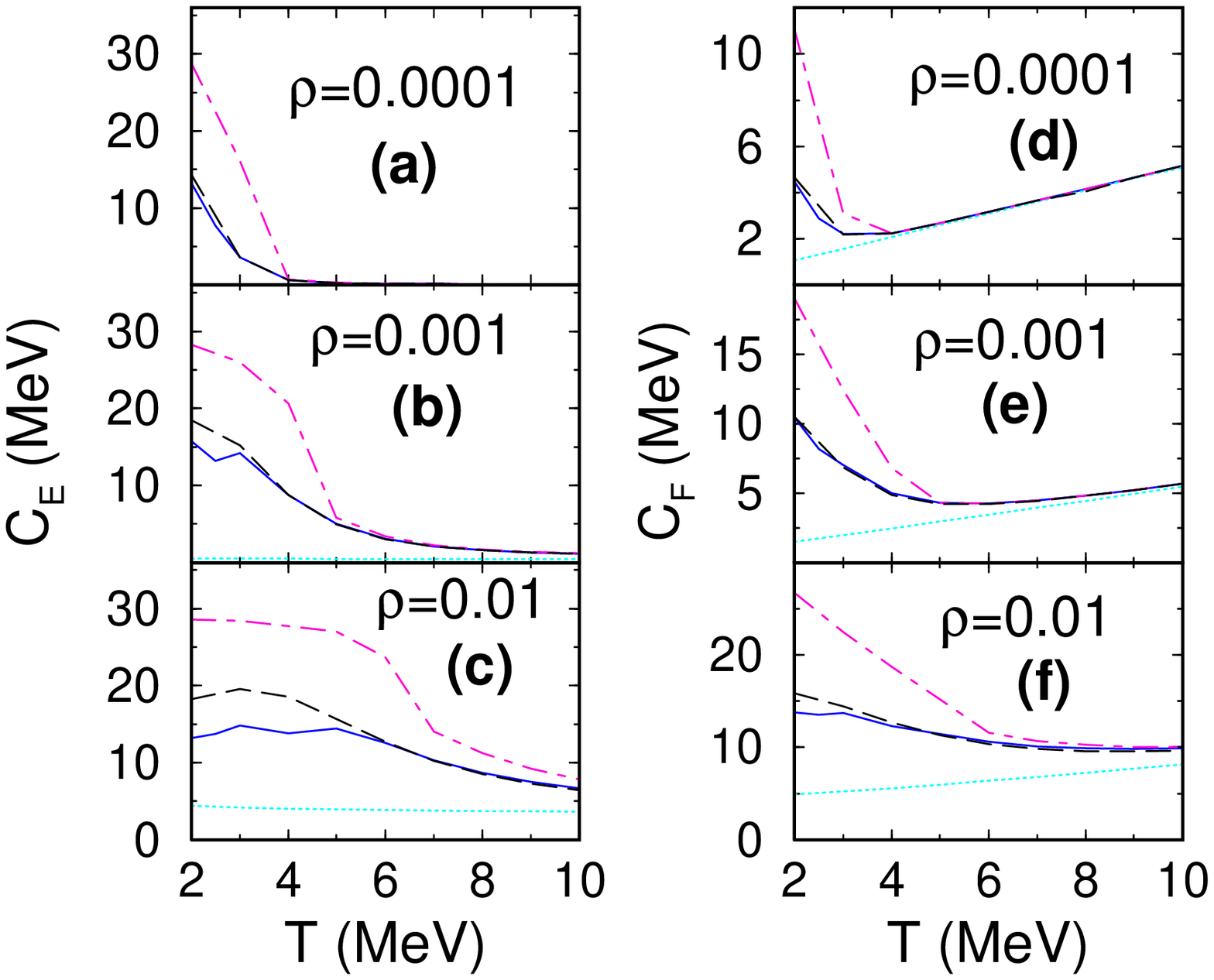}
\caption{ The symmetry coefficients $C_E$ and $C_F$ displayed as a
function of temperature at densities $\rho$=0.0001, 0.001, and
0.01 fm$^{-3}$. The notations are the same as described in 
the caption to Fig.~5.}
\end{figure}

\noindent
nucleons get bound in larger clusters; 
the pressure is also then lesser compared to the LSM model;
the deviations are significant, particularly
at higher density and at lower temperature.

The composition of matter at different densities and temperatures for
symmetric nuclear matter are shown in Fig.~2 through the charge multiplicities. 
The left, middle and right
panels correspond to temperatures $T$=2, 4, and 8 MeV, respectively,
at three different baryon densities
$\rho =$ 0.0001, 0.001 and 0.01 fm$^{-3}$.  The full lines
display the results from the SMF, the dashed 
lines are the ones from the LSM model
and the dot-dashed lines correspond to the ones from the
SNC model. Examination of
the results brings out a few important findings: i) At very
low densities, the multiplicities 
up to $Z=$2 are practically
the same in  the SMF and LSM models. In the SMF, heavier
fragments may be formed, but that is 
insignificant. ii) With
increase in density, heavy fragment 
formation can no longer
be neglected. Increase in temperature hinders the heavy cluster
formation. The multiplicity distributions display  a saw-toothed
nature. This is odd-even effect due to inclusion of pairing
in the phenomenological binding energies of the nuclear clusters;
this effect is diluted with increase in temperature.
iii) At relatively low temperatures and higher densities, in
the SNC model, the matter consists of only nucleons and very
heavy nuclei; the matter resembles liquid-like along with a
negligible fraction of nucleonic gas. 
These features are observed at
all the three densities at $T=$2 MeV and at $\rho =$0.01
fm$^{-3}$ for $T=$4 MeV. These results are not shown in the figure.
With increasing temperature and decreasing density, the liquid-like
structures disappear. These features appear from a delicate dependence
of the chemical potential on the density and temperature and the
dependence of fragment binding energy on increasing fragment mass which
tends to saturate at $\sim $16 MeV per nucleon.

 The evolution of symmetry free energy per baryon $f_{sym}$ as a 
function of asymmetry $X^2$ at different temperatures and densities
is displayed in Fig.~3. The symmetry free energy 
for a given density and temperature is defined as
\bea
f_{sym}~=~f(X)-f(0)+\frac{1}{\rho }
\sum_i[\rho_i(X)-\rho_i(0)]B_c^i,
\eea
where $B_c^i$ is the Coulomb contribution to the binding energy
of the $i$-th fragment species. The symmetry energy $e_{sym}$
can be defined likewise. The dashed, dot-dashed and the full lines
correspond to calculations at densities of $\rho =$0.0001, 0.001 and 0.01 
fm$^{-3}$, respectively. 
In Figs.~3(a), 3(b) and 3(c), the symmetry free energies per nucleon
in the LSM (thick black lines) and the SMF (thin
blue lines) are compared at the three densities at different
temperatures. As seen in Fig.~3(a) at $T=$2 MeV, at the lowest
density, the two calculations yield nearly
the same results; with increasing density, the difference in the two
model predictions shows up prominently because of the formation
of larger clusters in the SMF. This difference
washes out gradually with increasing temperature 
and the results are not discernible as seen in 
Fig.~3(b) and Fig.~3(c). In Fig.~3(d), the symmetry free energies
at $T=$2 MeV for the three densities from the SNC (thick magenta
lines) and MF (thin cyan lines) models are compared.
In the liquid-drop mass formula, the symmetry energy
is taken to be linear in $X^2$. This is seen to be nearly true also
for symmetry free energy of nuclear matter in a density region around the
saturation density \cite{xu} in the MF model;
we find the same  for dilute nuclear matter as is shown in Fig.~3(d). 
The symmetry energies can then be written as 
\bea
e_{sym}~=~C_E~X^2,~~~~~~f_{sym}~=~C_F~X^2,
\eea
where $C_E$ and $C_F$ are the symmetry energy and symmetry free
energy coefficients. Clusterization affects this linearity at
low temperatures; as the temperature is increased, the linearity 
tends to be restored as seen from Fig.~3(b) and Fig.~3(c). The symmetry
energy has a similar behavior and is not shown here. 

One interesting result that is borne out of the SMF calculation
is that at a given $T$ and $\rho$, asymmetric nuclear matter
may become more stable than symmetric matter [$f(X)<f(0)$].
This may result in negative $f_{sym}$ as is shown in Fig.~3(a).
This happens at relatively higher densities.
We find it to be due to the absence of
isospin-conjugate (mirror) nuclei because of Coulomb interaction.
Restricting the sum in Eq.~(3) to only mirror
nuclei removes this negativity. 
In the SNC model, clusters always occur in
isospin-conjugate pairs, $f_{sym}$ is then always positive as seen 
from Fig.~3(d).

The symmetry entropy per baryon, defined as $s_{sym}~=~s(X)-s(0)$
is presented as a function of asymmetry for different densities
and temperatures in Fig.~4. The notations used are the same as 
those used for Fig.~3. Comparison
between the MF and the SNC models at the three densities is displayed
in Fig.~4(b) and Fig.~4(d). In 
the MF model, the symmetry entropy decreases with asymmetry.
This is akin to the {\it entropy of mixing }. The entropy per
nucleon of an ideal two-component nucleon gas is given, 
from Gibbs-Duhem relation, by
\bea
s(X)~=~\frac{5}{2}-\frac{\rho_n}{\rho} \ln \zeta_n
-\frac{\rho_p}{\rho} \ln \zeta_p .
\eea
Using the fact that for low density nucleonic matter
\bea
\rho_{n,p}\simeq \frac {2}{\lambda^3}\zeta_{n,p},
\eea
the symmetry entropy can be shown to behave as
\bea
s_{sym}(X)~=~-\frac{1}{2}X^2\left (1+ \frac{1}{6}X^2+....\right ),
\eea
which for low values of asymmetry  decreases linearly with $X^2$.
This is independent of both temperature and density and  is nearly
manifested for the low density nuclear matter we have considered.
As can be seen from the comparison with the SNC calculations,
clusterization changes this behavior; in the density and temperature 
domain where clusterization becomes important, the symmetry entropy
is larger compared to that in the MF model and can even be positive. 
In Fig.~4(a) and Fig.~4(c), comparison is made between 
the results from LSM and SMF. At
high temperatures and low densities, the difference in the results
from these models is not discernible; however, at relatively high density
and low temperature, $s_{sym}$ in the LSM  is appreciably
larger than that in the SMF as displayed in Fig.~4(a) 
for $\rho =$0.01 fm$^{-3}$ and $T=$ 4 MeV. This is attributed to the 
relatively rapid growth of multiplicity (mostly of neutrons
at the cost of $p$, $He^3$ and $\alpha $) in LSM model 
compared to that in SMF.

The conventional definition of the symmetry coefficients $C_E$ and $C_F$
as given by Eq.~(17) holds good when $e_{sym}$ and $f_{sym}$ are
linear or nearly linear in $X^2$. In Fig.~5, we display these symmetry
coefficients as a function of density for temperatures when the 
symmetry energies are nearly linear in $X^2$ in the different
models.  In the density region investigated,
in the MF model, the symmetry coefficients increase linearly
with density in contrast to a power-law dependence 
$\sim (\rho/\rho_0)^\gamma$
with $\gamma \sim $0.7  at relatively higher densities \cite{sam1}.
The symmetry coefficients are enhanced significantly when clusterization
is considered. This is most prominent in the SNC model
(dot-dashed lines), where as already
stated, at lower temperatures and higher densities, the system
is more liquid-like; this is reflected in the value of $C_E$
in Fig.~5(a), which at higher densities saturates to $\sim$28 MeV,
the value for normal nuclear matter as given in the liquid-drop
mass formula we use \cite{mye1}. 
The symmetry coefficient $C_E $ in the LSM
and in SMF are close at high temperature $T$= 8 MeV,
but differ at $T$= 4 MeV where clusterization is more 
important. This difference is filled up for symmetry free
energy coefficient $C_F$ due to the symmetry free entropy
[shown in Fig.~4(a)] resulting in practically the same
$C_F$ in both the models.

 The temperature dependence of the symmetry coefficients $C_E$
and $C_F$ at a few densities in all the models considered are displayed
in the left and right panels of Fig.~6, respectively. In the
mean-field model, shown by the dotted lines, $C_E$ is linear 
and approximately constant for a given density as seen earlier
\cite{sam1}. At very low density $\rho $=0.0001 fm$^{-3}$,
it is nearly zero and not discernible in the figure. In the
condensation models
a pronounced increase in the values of $C_E$
is observed, particularly at low temperatures. In the SNC model
this value is close to that for normal nuclear matter as expected. 
At larger
temperatures, values of $C_E$ obtained in the condensation models
approach those calculated from MF.

The symmetry free energy coefficient $C_F$, by 
our choice, can be written as
\bea
C_F~=~C_E-T \frac{s_{sym}}{X^2} .
\eea
For dilute nuclear matter, in the MF model, $C_E$ is linear
in $T$ at a particular density and $s_{sym}$, as stated
earlier, is negative and proportional to $X^2$. A linear
increase of $C_F$ with temperature is therefore expected
in the mean-field model which is realized as displayed by the
dotted lines in the right panels of Fig.~6. At higher temperatures,
results for $C_F$ in all the models tend to merge, particularly
at low density because of the hindrance to form clusters. At
lower temperatures, $C_F$ is significantly higher with clusterization
compared to that in the MF.
In Eq.(21), the first term on the right hand side decreases
with temperature whereas the second term increases with condensation;
this interplay causes a minimum in $C_F$ which is more
pronounced at lower densities.

\section{Concluding Remarks}

The role of condensation on some properties of dilute nuclear matter,
namely, the nuclear EOS, the symmetry free energy, entropy and the
symmetry coefficients has been addressed in this article in the 
$S$-matrix framework. This approach has the advantage that the
relevant observables can be directly connected to the experimentally
measured quantities like the nuclear binding energies and the scattering
phase-shifts. This approach
contains no interaction potential parameters and hence, the results are
mostly model-independent.

Except at very low densities, the nuclear EOS in the $S$-matrix approach
differs appreciably from the one obtained in the MF model. The 
MF model, supplemented with Maxwell's construction
displays a liquid-gas phase transition on isothermal
compression, in the $S$-matrix approach, the system responds to
the compression by a marked growth of clusters out of the 
dilute nucleonic gas. The growth is hindered with isochoric
heating.

One of the very remarkable features of the results on symmetry
energies (both $e_{sym}$ and $f_{sym}$) in SMF is that the
symmetry energies are generally nonlinear in $X^2$,
contrary to the results in the MF model. The symmetry energies 
may even be negative in the SMF at low temperatures, 
high densities and at small $X$. 
The symmetry entropy
similarly displays a subtle behavior with clusterization.
 In the conventional definition, at a particular temperature
and density, the symmetry coefficients are then no longer independent of
the asymmetry parameter and may be negative. In the region where the
symmetry energies are practically linear in $X^2$, the symmetry energy
and free energy coefficients in the $S$-matrix approach are found to
be appreciably larger compared to those obtained in the MF model. The
nuclear EOS and the symmetry energies and coefficients are thus seen to
have a significant dependence on the many-nucleon correlations or
cluster formation in the low density nuclear matter.
Inclusion of these effects are thus warranted for the study of
physical phenomena like supernova dynamics that depend sensitively
on the symmetry energy and its temperature and density dependence.

\acknowledgments{The authors acknowledge the support from the Department
of Science  \& Technology, Government of India.}

\end{document}